\documentclass[aps,prb,twocolumn,superscriptaddress,longbibliography,footinbib]{revtex4-1}
\usepackage{graphicx}
\usepackage{latexsym}
\usepackage{amssymb}
\usepackage{amsmath}
\usepackage{amsfonts}
\usepackage{upgreek}
\usepackage{float}
\usepackage{bm}
\usepackage{multirow}
\usepackage{color}
\usepackage[T1]{fontenc}
\usepackage{hyperref}
\usepackage{subfigure}

\hypersetup{
colorlinks = true,
linkcolor = [rgb]{0.70,0.13,0.13},
citecolor = [rgb]{0.13,0.55,0.13},
urlcolor  = [rgb]{0.25, 0.41, 0.88}}
\newcommand{\ket}[1]{|#1\rangle}
\newcommand{\bra}[1]{\langle#1|}

\begin{document}
\title{Dynamical scaling laws in the quantum $q$-state clock chain}
\author{Jia-Chen Tang}
\affiliation{College of Physics, Nanjing University of Aeronautics and Astronautics, Nanjing, 211106, China}
\affiliation{Key Laboratory of Aerospace Information Materials and Physics (Nanjing University of Aeronautics and Astronautics), MIIT, Nanjing 211106, China}

\author{Wen-Long You}
\affiliation{College of Physics, Nanjing University of Aeronautics and Astronautics, Nanjing, 211106, China}
\affiliation{Key Laboratory of Aerospace Information Materials and Physics (Nanjing University of Aeronautics and Astronautics), MIIT, Nanjing 211106, China}

\author{Myung-Joong Hwang}
\affiliation{Division of Natural and Applied Sciences, Duke Kunshan University, Kunshan, Jiangsu 215300, China}
\affiliation{Zu Chongzhi Center for Mathematics and Computational Science, Duke Kunshan University, Kunshan, Jiangsu 215300, China}

\author{Gaoyong Sun}
\thanks{Corresponding author: gysun@nuaa.edu.cn}
\affiliation{College of Physics, Nanjing University of Aeronautics and Astronautics, Nanjing, 211106, China}
\affiliation{Key Laboratory of Aerospace Information Materials and Physics (Nanjing University of Aeronautics and Astronautics), MIIT, Nanjing 211106, China}

\begin{abstract}
We show that phase transitions in the quantum $q$-state clock model for $q \leq 4$ can be characterized by an enhanced decay behavior of the Loschmidt echo via a small quench. 
The quantum criticality of the quantum $q$-state clock model is numerically investigated by the finite-size scaling of the first minimum of the Loschmidt echo and the short-time average of the rate function. 
The equilibrium correlation-length critical exponents are obtained from the scaling laws which are consistent with previous results. 
Furthermore, we study dynamical quantum phase transitions by analyzing the Loschmidt echo and the order parameter for any $q$ upon a big quench. 
For $q \leq 4$, we show that dynamical quantum phase transitions can be described by the Loschmidt echo and the zeros of the order parameter.
In particular, we find the rate function increases logarithmically with $q$ at the first critical time.
However, for $q > 4$, we find that the correspondence between the singularities of the Loschmidt echo and the zeros of the order parameter no longer exists.
Instead, we find that the Loschmidt echo near its first minimum converges, while the order parameter at its first zero increases linearly with $q$.
\end{abstract}

\maketitle

\section{Introduction}
Continuous phase transitions in equilibrium are central concepts in quantum many-body systems \cite{Sachdev1999}. 
The nature of phase transitions can usually be charaterized by the universality classes and the order parameters 
from the renormalization group \cite{Wilson1974,Wilson1975} and the finite-size scaling theory \cite{Fisher1972,Fisher1974}. 
Using the theoretical tools of quantum information science, quantum phase transitions and critical phenomena in equilibrium can also be probed by the quantum entanglement \cite{osterloh2002scaling,horodecki2009quantum,eisert2010colloquium}, 
the ground-state fidelity \cite{You2007,Venuti2007,Chen2008,Gu2008,Yang2008,Gu2010,Sun2017,Zhu2018,Lu2018,sun2022biorthogonal,wang2022quantum}
and the Loschmidt echo \cite{Quan2006decay,Hwang2019Universality,tang2022dynamical}. In contrast to the quantum entanglement and the ground-state fidelity, which are properties of the ground state, 
the Loschmidt echo is a non-equilibrium quantity and is much easier to be measured in experiments upon a sudden quench.
Recently, the dynamical scaling laws of the Loschmidt echo are established to extract equilibrium critical exponents of many-body systems by the finite-size scaling theory 
for second-order phase transitions \cite{Hwang2019Universality}. 
For instance, the universality class of phase transitions in one-dimensional Hermitian \cite{Hwang2019Universality} and non-Hermitian transverse field Ising chain \cite{tang2022dynamical}
were identified by the Loschmidt echoes.

On the other hand, the generalization of phase transitions to nonequilibrium systems \cite{odor2004universality,weimer2021simulation} is attractive 
from the perspective of exploring unconventional phase transitions.
Recently, an interesting nonequilibrium phase transition, named the dynamical quantum phase transition (DQPT) \cite{heyl2013dynamical,heyl2018dynamical,marino2022dynamical}, 
is found to occur during the real-time evolution of a system upon a sudden quench.
DQPTs take place after a big sudden quench of the system across equilibrium quantum critical points in the thermodynamic limit \cite{heyl2013dynamical,hagymasi2019dynamical,sun2020dynamical}, 
which have been investigated in various systems \cite{jurcevic2017direct,flaschner2018observation,xu2020measuring,wang2019simulating,tian2019observation,guo2019observation,nie2020experimental,tian2020observation,wu2022dynamical}.
The DQPT arising from the large quench is often characterized by the Loschmidt echo singularities \cite{heyl2013dynamical} 
and the zeros of an order parameter \cite{heyl2013dynamical,hagymasi2019dynamical,sun2020dynamical,jurcevic2017direct} at the critical time $t_c$.
It is argued recently that the dynamics of such a DQPT is analogous to a two-level system dynamics \cite{zakrzewski2022dynamical}. Whether the DQPT can exhibit a complex many-body dynamics or merely behave as a two-level system remains to be understood \cite{van2022anatomy,corps2022theory,kuliashov2022dynamical}.

Motivated by the use of the Loschmidt echo both to characterize equilibrium phase transitions and to unveil novel non-equilibrium phase transitions, 
we investigate the quench dynamics in the quantum $q$-state clock model, where Loschmidt echoes can be analytically solved \cite{karrasch2017dynamical,wu2020nonequilibrium}
for some special quench protocols. In particular, we focus on the following two questions: 
First, whether the Loschmidt echo can be used to detect equilibrium second-order phase transitions in the presence of discrete symmetries higher than the $Z_{2}$ symmetry.
Second, what is the relationship between the singularities of the Loschmidt echo and the zeros of an order parameter of the DQPTs in the $q$-state clock model.

In this paper, we first use the Loschmidt echo to explore equilibrium second-order phase transitions in the $q$-state clock model and to identify the finite-size dynamical scaling 
and the correlation-length critical exponents for $q \leq 4$. 
We find that the decay of the Loschmidt echo is enhanced near the equilibrium critical point, through which we obtain the equilibrium correlation-length critical exponents. 
We show that in the absence of the knowledge of the critical point and the phase transition, 
one can use the short-time average rate function to study the quantum criticality even in the presence of $Z_q$ symmetries.
In addition, we develop a numerically efficient method to perform the finite-size scaling for extracting equilibrium critical exponents by simply using the first minima of the Loschmidt echoes.
Furthermore, we investigate the DQPTs of the $q$-state clock model upon a big quench. 
We derive analytical solutions for both the Loschmidt echo and the order parameter for arbitrary $q$, 
with which we show that DQPTs arising from the Loschmidt echo singularity can be connected to the zeros of the order parameters for $q \leq 4$.
In particular, we find the value of the rate function of the Loschmidt echo of $q=4$ is twice as much as that of $q=2$. 
In addition, we show that the rate function increases logarithmically with $q$ at the first critical time $t_{c}$ in the regime $q \leq 4$.
In contrast, for $q > 4$, we find that the Loschmidt echo singularity does not correspond to the zeros of the order parameters.
The Loschmidt echo at its first minimum converges, and the order parameter at its first zero increases linearly with $q$.

This paper is organized as follows. 
In Sec.\ref{sec:qclockmodel}, we introduce the quantum $q$-state clock model. 
In Sec.\ref{sec:LE}, we discuss the scaling law of the Loschmidt echo.
In Sec.\ref{sec:smallquench}, we study a small quench dynamics for $q \leq 4$ in the vicinity of the equilibrium quantum critical point of the $q$-state clock model and extract equilibrium critical exponents.
In Sec.\ref{sec:bigquench}, we study the DQPTs of the $q$-state clock model upon a big quench for any $q$ and its relation to the zeros of order parameters. 
In Sec.\ref{sec:conclusion}, we summarize our results.

\section{q-state clock model}
\label{sec:qclockmodel}
We consider a general one-dimensional quantum $q$-state clock model of $N$ sites with periodic boundary condition, whose Hamiltonian is given by \cite{matsuo2006berezinskii,ortiz2012dualities,chen2017phase,sun2019phase},
\begin{align}
H = -J \sum_{j=1}^{N} (U_{j+1}^{\dag}U_{j} + U_{j}^{\dag}U_{j+1} ) - h \sum_{j=1}^{N}(V_{j} + V_{j}^{\dag}),
\label{Eq:clock}
\end{align}
where the kinetic energy term is represented by the unitary operator $U_{j}$ with the coupling coefficient $J$, 
the other unitary operator $V_{j}$ denotes the potential energy term with the coupling constant $h$. The periodic boundary condition is imposed as $U_{N+1} = U_{1}$.
We label the $q$ states of the local Hilbert space at the site $j$ by $|0\rangle_{j},...,|l \rangle_{j},...,|q-1 \rangle_{j}$ with $0<l<q-1$. 
In these orthogonal bases, the operators $U_{j}$ and $V_{j}$ are written as,
\begin{align}
U_j =
\left( \begin{array}{cccccc}
1 & 0 & 0 & 0 & \ldots & 0 \\
0 & \omega & 0 & 0 & \ldots & 0 \\
0 & 0 & \omega^{2} & 0 & \ldots & 0 \\
0 & 0 & 0 & \omega^{3} & \ldots & 0 \\
\vdots & \vdots & \vdots & \vdots & \ddots & \vdots \\
0 & 0 & 0 & 0 & \ldots & \omega^{q-1}
\end{array} \right),
\end{align}
\begin{align}
V_j =
\left( \begin{array}{cccccc}
0 & 1 & 0 & 0 & \ldots & 0 \\
0 & 0 & 1 & 0 & \ldots & 0 \\
0 & 0 & 0 & 1 & \ldots & 0 \\
\vdots & \vdots & \vdots & \vdots & \ddots & \vdots \\
0 & 0 & 0 & 0 & \ldots & 1 \\
1 & 0 & 0 & 0 & \ldots & 0
\end{array} \right).
\end{align}
Here, the unitary operator $U_{j}$ is a diagonal matrix in which diagonal elements are $\omega^{k}$ with $\omega = e^{i 2\pi /q} \equiv e^{i \theta}$ and $k=0, 1 ,... ,q-1$ for arbitrary integers $q$. 
$U_j$ and $V_j$ obey the following relations,
\begin{align}
V_{j}U_{j} =&{} \omega U_{j}V_{j}, \\ 
U_{j}^{q} =&{} V_{j}^{q}=1.
\end{align}
The quantum $q$-state clock model which has the $Z_q$ symmetry undergoes second-order phase transitions \cite{matsuo2006berezinskii,ortiz2012dualities,chen2017phase,sun2019phase}
for $q \leq 4$ and Berezinskii-Kosterlitz-Thouless (BKT) transitions for $q > 4$. In the following, we will study the $q$-state clock model out of equilibrium by considering the following two cases: 
One case is that we will discover equilibrium second-order phase transitions of $q \leq 4$ by the Loschmidt echo as well as its short-time average rate function upon a small sudden quench. 
The other case is that we will investigate DQPTs by the Loschmidt echo and the order parameter upon a big sudden quench for any $q$.

\begin{figure}[tb]
\includegraphics[width=8.7cm]{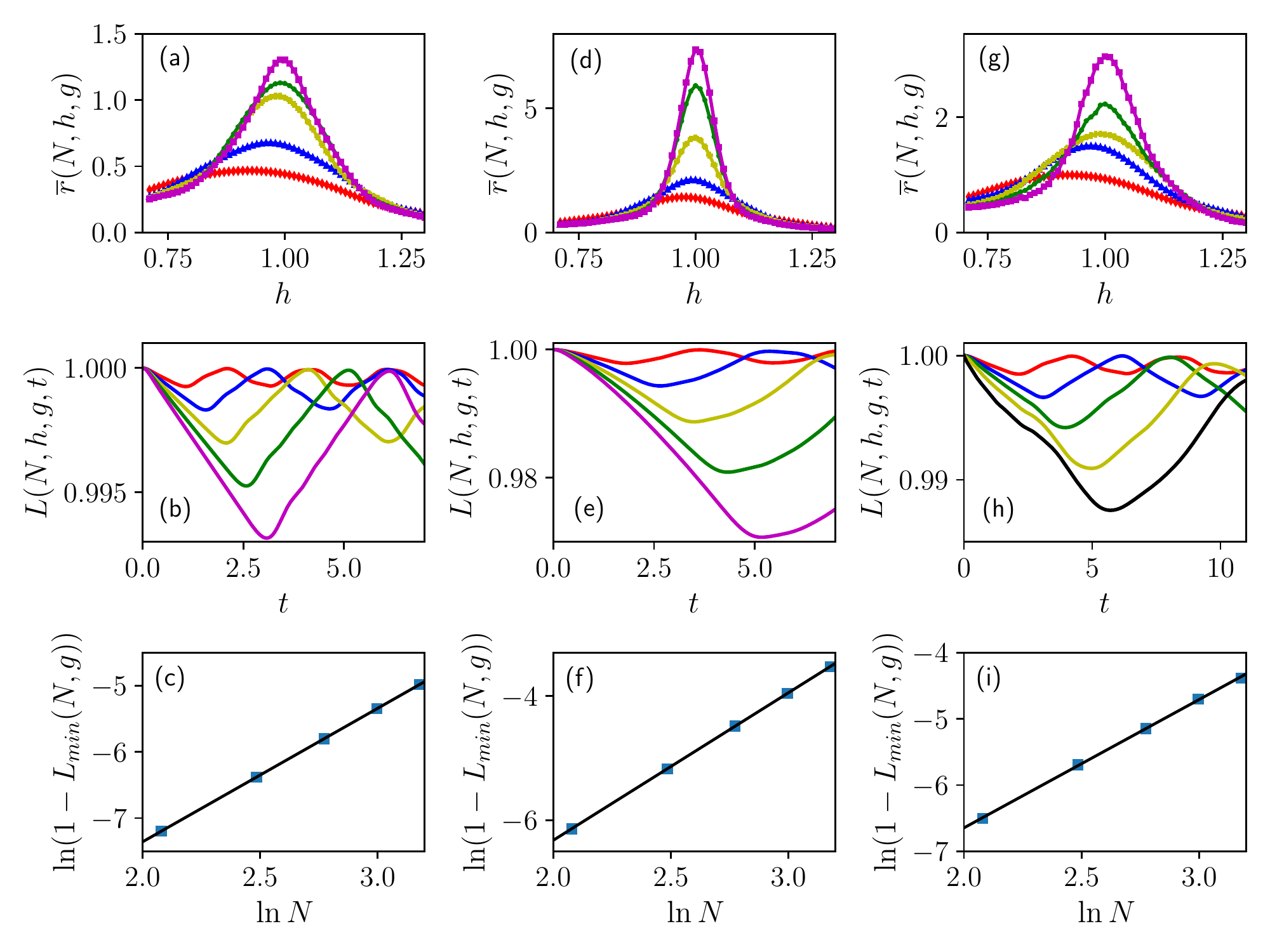}
\caption{ Scaling of the short-time average rate function and the Loschmidt echo of the $q$-state clock model.
(a) The short-time average rate function $\bar{r}(N,h,g)$ with respect to $h$ with $g=0.01$ in different lattice sizes $N =8,12,16,20,24$ (from bottom to top along the peaks) for $q=2$.
(b) The Loschmidt echo $L(N,h,g,t)$ at the peak position $h_{\ast}$ of $\bar{r}(N,h,g)$ in (a) with $g=0.01$ as a function of time $t$ for lattice sizes $N =8,12,16,20,24$ (from top to bottom along the first minima).
(c) Finite-size scaling of $1-L_{min}(N,g)$ obtained from (b) as a function of lattice sizes $N$, where the blue square symbols are numerical values, the black solid line denotes the fitting curve. 
The correlation-length critical exponent $\nu=0.992$ is obtained from the fitting curve. 
Here (d), (e) and (f) represent the short-time average rate function, the Loschmidt echo and the finite-size scaling of $1-L_{min}(N,g)$ for $q=3$, respectively.
The corresponding data for $q=4$ are shown in (g), (h) and (i), respectively. 
The correlation-length critical exponents obtained from fitting curves are $\nu=0.842$ for $q=3$ and $\nu=1.033$ for $q=4$.}
\label{ratefigp2}
\end{figure}


\section{Loschmidt echo}
\label{sec:LE}
In this section, we briefly introduce the Loschmidt echo and its rate function.
Given an arbitrary initial state $\ket{\psi_{0}}$, the time evolution under a post-quenched time-independent Hamiltonian $H_{f}$ is given by,
\begin{align}
\ket{\psi(t)} = e^{-iH_{f}t} \ket{\psi_{0}},
\end{align}
where $\hbar =1$.
The Loschmidt echo is defined by the return probability (or time-evolved fidelity),
\begin{align}
L(t) = |\langle \psi_{0}| e^{-iH_{f}t} | \psi_{0} \rangle|^2, 
\label{Eq:Lt}
\end{align}
with the Loschmidt amplitude $G(t) = \langle \psi_{0} | \psi(t) \rangle$.

It has been shown that the decay of the Loschmidt echo can be enhanced by the equilibrium quantum criticality\cite{Quan2006decay}. 
The first minimum of the Loschmidt echo at the time $t_{min,1}$ is recently shown to scale as \cite{Hwang2019Universality},  
\begin{align}
1-L_{min}(N,g) \propto g^2 N^{2/\nu},
\label{Eq:LEmin}
\end{align}
at the equilibrium critical point for second-order phase transitions. Here $g$ is the small constant step defined by,
\begin{align}
g = h_{f}-h_{i},
\end{align}
with $h_{i}$ and $h_{f}$ are two coupling constants to control quench protocols.
The dynamical scaling law in Eq.(\ref{Eq:LEmin}) that governs the critically enhanced decay behavior of the Loschmidt echo with respect to $N$
can be used to extract the equilibrium correlation-length critical exponent $\nu$.

We note that the Loschmidt echo $L_{min}(N,g)$ exhibits the scaling law in Eq.(\ref{Eq:LEmin}) for a small quench in the vicinity of the equilibrium critical point $h_c$. 
This poses a challenge for using the Loschmidt echo to diagnose the equilibrium criticality if a prior knowledge about the precise value of the critical point $h_c$ is absent \cite{Hwang2019Universality,tang2022dynamical}. 
In the following section, we propose to use a short-time average of the rate function \cite{tang2022dynamical},
\begin{align}
\bar{r}(N,h,g)=-\frac{1}{N}\frac{\ln(\bar{L}(N,h,g))}{g^2},
\label{rateAver}
\end{align}
which is analogous to the ground-state fidelity susceptibility to find $L_{min}(N,g)$.
Here $\bar{L}(N,h)$ is the short-time average of the Loschmidt echo within the time duration $T$ defined by,
\begin{align}
\bar{L}(N,h,g)=\frac{1}{T} \int_{0}^{T} L(N,h,g,t)dt.
\label{LEAver}
\end{align}
We note that the time duration $T$ for the average should exceed $t_{min,1}$ in order to recover the value of  $L_{min}(N,g)$.

On the other hand, the rate function of the Loschmidt echo given by,
\begin{align}
r(t)=-\frac{1}{N} \ln L(t)
\label{Eq:RF}
\end{align}
can reveal Loschmidt echo singularities at the critical time $t_c$ upon a big sudden quench \cite{heyl2013dynamical,heyl2018dynamical,marino2022dynamical}. 
The singularities of the rate function indicate that a system undergoes DQPTs. 
In the following, we will use the rate function in Eq.(\ref{Eq:RF}) to analyze the DQPTs.

\begin{figure}[tb]
\includegraphics[width=8.1cm]{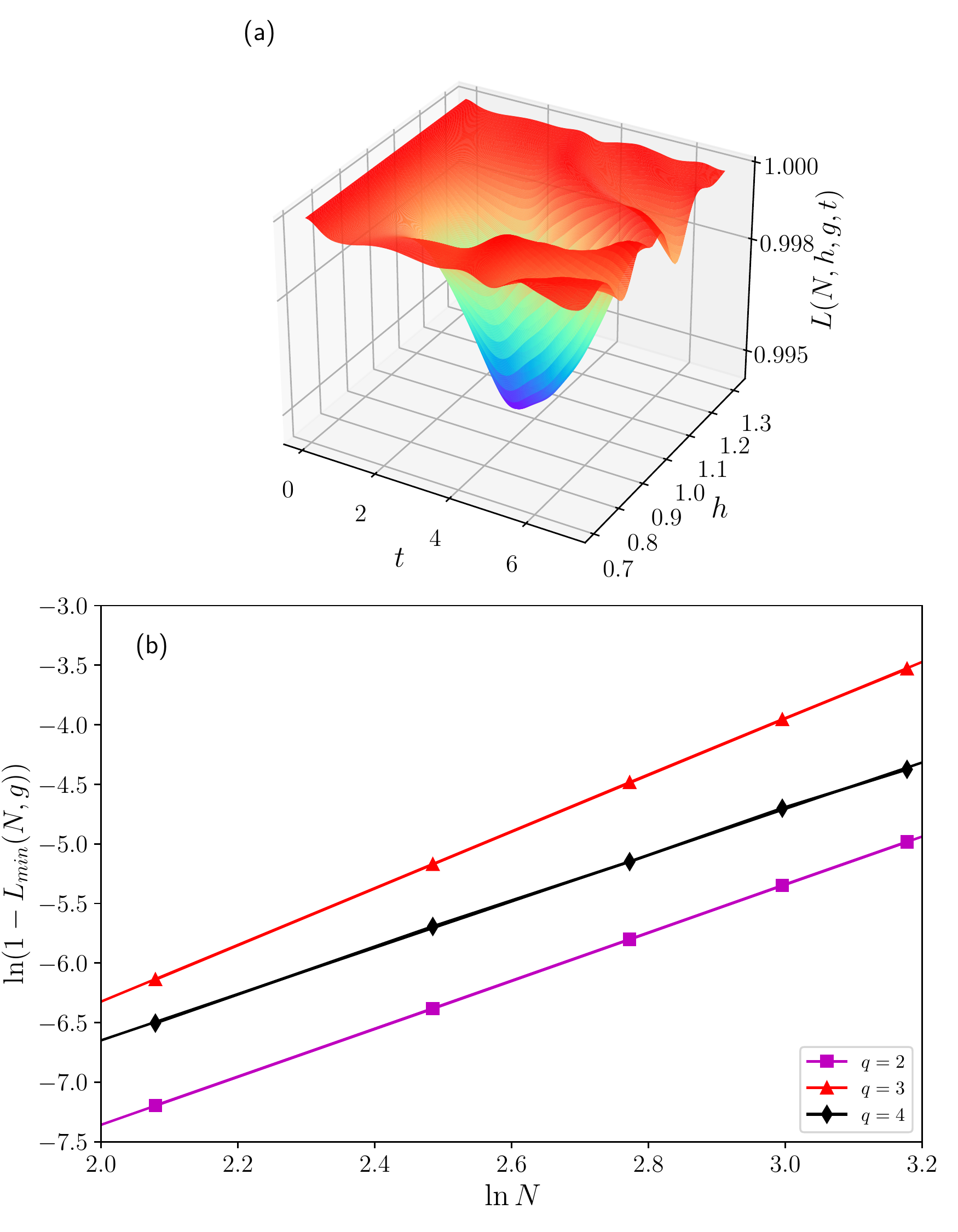}
\caption{ Evolution and the scaling of the Loschmidt echo.  
(a) Time evolution of the Loschmidt echo $L(N,h,g,t)$ as the function of $h$ and $t$ for $q=3$ with $g=0.01$ in $N = 16$ lattice sites, which exhibits a decay behavior enhanced by the quantum criticality.
(b) Finite-size scaling of the minima of the Loschmidt echoes $L_{min}(N,g)$ for $q = 2$ (purple square),  $q = 3$ (red triangle), $q = 4$ (black diamond) with sizes $N = 8, 12, 16, 20, 24$.
The correlation-length critical exponents obtained from fitting curves are $\nu=0.992$ for $q=2$, $\nu=0.842$ for $q=3$ and $\nu=1.029$ for $q=4$.}
\label{ratefigp34}
\end{figure}


\section {Dynamics upon small quench}
\label{sec:smallquench}
In this section, we will study the dynamics of the $q$-state clock model in Eq.(\ref{Eq:clock}) upon a small quench. 
First of all, we obtain the ground-state $\ket{\psi_{0}}$ of the Hamiltonian in Eq.(\ref{Eq:clock}) at the coupling $h_{i}$, then compute the Loschmidt echo using Eq.(\ref{Eq:Lt}) 
for a quench protocol where the coupling constant is suddenly changed from the initial $h_{i}$ to a final $h_{f}$ with a small constant step $g=0.01$. 
We calculate the time-evolved wave function $\ket{\psi(t)}$ using the time-dependent density matrix renormalization group (t-DMRG) \cite{schollwock2011density,orus2014practical,daley2004time,vidal2004efficient,vidal2007classical}
with a time step $\Delta t = 10^{-3}$ under periodic boundary conditions \footnote{The errors in the t-DMRG simulations come from the truncation $\epsilon_1=\displaystyle\sum_{\alpha=M+1}^{\chi}(\lambda_{\alpha})^2$ and the Trotter expansion $\epsilon_2 \propto (\Delta t)^{2p} T^{2}$ [\onlinecite{vidal2004efficient}], where $\lambda_{\alpha}$ are the singular values of the wave function $\ket{\psi(t)}$, $\chi$ is the total number of singular values, $M$ is the number of singular values (states) kept, $p$ is the order of the Trotter expansion and $T$ is the total time duration. We use the second-order Trotter expansion $p=2$ by keeping $M=200$ states.}, where we choose $J=1$. 
In this following, we will focus on the cases of $q \leq 4$
as the dynamical scaling laws of the Loschmidt echo given in Eq.(\ref{Eq:LEmin}) are argued to be valid for second-order phase transitions \cite{Hwang2019Universality}. 
The dynamical scaling laws of the Loschmidt echo are poorly understood for equilibrium BKT transitions to the best of our knowledge, which are left for future study. 
We also note that the numerical simulations for $q > 4$ are difficult using the t-DMRG method with periodic boundary conditions.

In the absence of any prior knowledge about the exact critical value of a model, we propose to use short-time average rate function to probe second-order phase transitions. 
Let us briefly summarize the procedure here and apply it to the $q$-state clock model.
We first calculate the short-time average rate functions $\bar{r}(N,h,g)$ from Eq.(\ref{rateAver}) by varying the coupling $h$ and find the pseudo critical points $h_{\ast}$, 
which are derived from the peaks of short-time average rate functions for each lattice $N$ as shown in Fig.\ref{ratefigp2}(a). 
We then perform numerical simulations with the t-DMRG upon a quench from this pseudo critical point $h_{\ast}$ to $h_{f} = h_{\ast} + g$ for $N = 8, 12, 16, 20, 24$ sites. 
The results of the Loschmidt echoes $L(N,h,g,t)$ presented in Fig.\ref{ratefigp2}(b) exhibit a decay and revival dynamics. 
The first minima of the Loschmidt echoes $L_{min}(N,g)$ are plotted in Fig.\ref{ratefigp2}(c) with respect to the lattice size $N$.  
According to the scaling law in Eq.(\ref{Eq:LEmin}), we obtain the critical exponent $\nu=0.991 \pm 0.003$ for $q=2$. 
Similarly, we find the critical exponent $\nu=0.844 \pm 0.006$ when $q=3$ and the critical exponent $\nu=1.038 \pm 0.022$ when $q=4$ as demonstrated in Fig.\ref{ratefigp2} from (d) to (i) \footnote{ We note that the critical exponents $\nu=(\nu_{max} + \nu_{min})/2 \pm (\nu_{max} - \nu_{min})/2$, where $\nu_{max}$ and $\nu_{min}$ are the maximal and minimal values of the critical exponents obtained by fitting the data $1-L_{min}$ from the combinations of three ($C_{5}^{3}$), four ($C_{5}^{4}$) and five ($C_{5}^{5}$) elements of the set $N=8,12,16,20,24$ sites using Eq.(\ref{Eq:LEmin})}. 

The above results of the critical exponents are consistent with the exact values \cite{sun2019phase}, which are $\nu=1$ for $q=2,4$ and $\nu=5/6$ for $q=3$. 
This demonstrates that the short-time average rate function 
is a valid method to probe phase transitions of the $q$-state clock model without knowing the critical values in advance. 
A drawback of this method, however, is that one has to choose a time duration for the average that may affect the precision on the retrieved values of critical exponents. 
This can be circumvented by instead using the first minima in the three-dimensional (3D) plot of the Loschmidt echoes [c.f Fig.(\ref{ratefigp34})(a)] as a probe to perform the finite-size scaling 
with Eq.(\ref{Eq:LEmin}) to extract equilibrium critical exponents.
The correlation-length critical exponents obtained from first minima of the Loschmidt echoes $L_{min}(N,g)$ 
are $\nu=0.991 \pm 0.003$ for $q=2$, $\nu=0.844 \pm 0.006$ for $q=3$ and $\nu=1.034 \pm 0.017$ for $q=4$, respectively [c.f Fig.(\ref{ratefigp34})(a)].

\begin{figure}[tb]
\includegraphics[width=8.3cm]{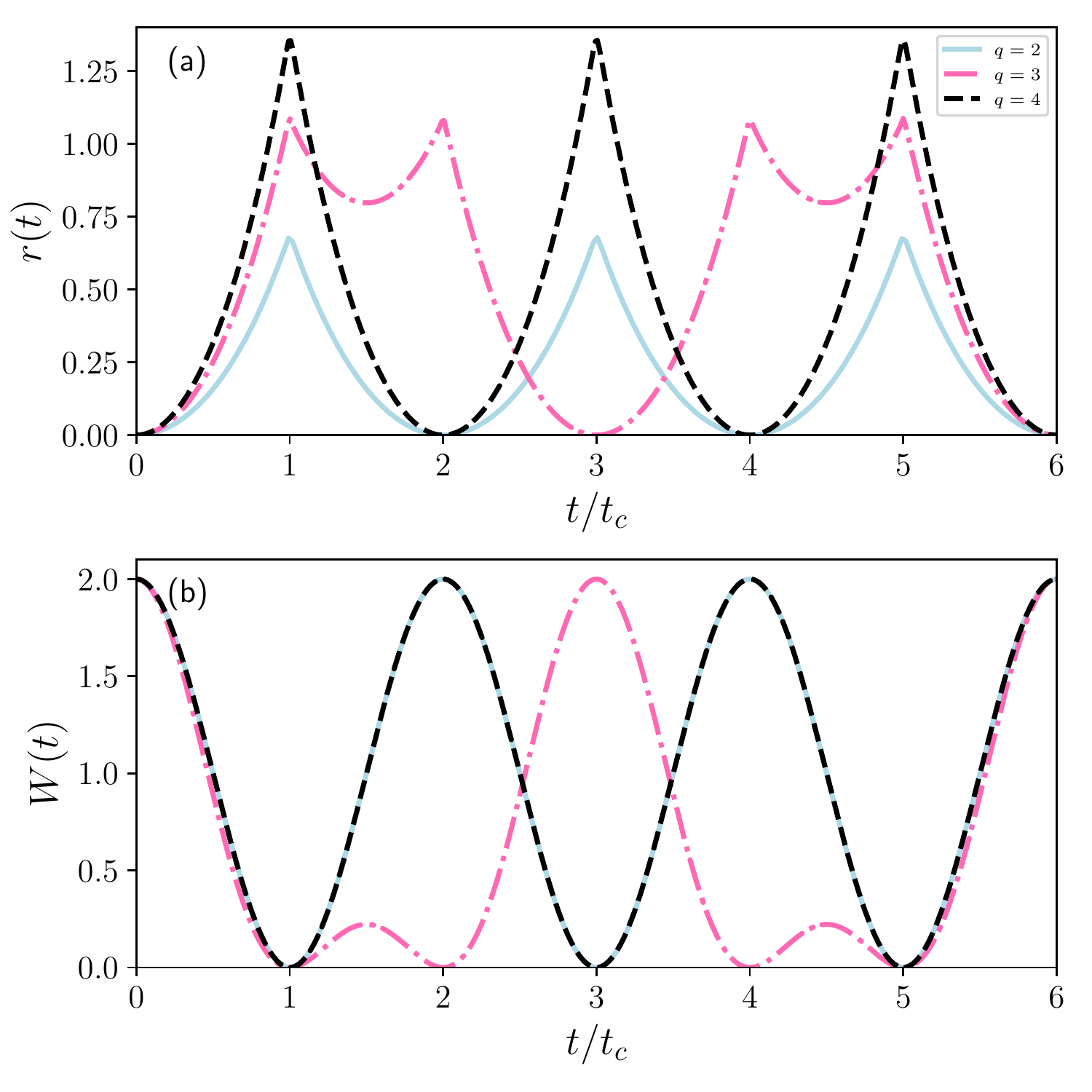}
\caption{Rate functions of Loschmidt echoes and order parameters upon a quench from $h_{i} = \infty$ to $h_{f}=0$.  
(a) The rate function $r(t)$ as the function of $t/t_{c}$ for $q \leq 4$ for $J=1$ and $N=100$ lattice sites, where the rate function $r(t)$ for $q=4$ is twice the value of that for $q=2$.
Here, the first critical times $t_{c}$ are $\pi/8$ for $q=2$, $2\pi/9$ for $q=3$ and $\pi/4$ for $q=4$, respectively. 
(b) The order parameter $W(t)$ with respect to $t_c$ with the same parameters as (a), whose zeros correspond to critical times $h_c$ shown in (a).}
\label{rateorderp234}
\end{figure}

\section {DQPT upon big quench}
\label{sec:bigquench}
Let us now consider the dynamics of the $q$-state clock model in Eq.(\ref{Eq:clock}) upon a big sudden quench.
We first consider the case by quenching the system from $h_{i}=\infty$ to $h_{f}=0$, in which we derive analytic solutions of the Loschmidt echoes as well as the order parameters for arbitrary $q$.
The results are separately discussed in two parts ($q \leq 4$ and $q > 4$) according to the universality classes of the model. 

The ground-state $\ket{\psi_{0}}$ of the $q$-state clock model in Eq.(\ref{Eq:clock}) at $h_{i}=\infty$ is the product state given by,
\begin{align}
\ket{\psi_{0}} = \bigotimes_{j=1}^{N} \ket{\phi}_{j},
\end{align}
where the local eigenstate $\ket{\phi}_{j}$ is
\begin{align}
\ket{\phi}_{j} = \left(\frac{1}{\sqrt{q}} \sum_{n=0}^{q-1} \ket{n} \right)_{j} 
=\frac{1}{\sqrt{q}}
\left( \begin{array}{ccccc}
1\\
1\\
\vdots\\
1 \\
1
\end{array} \right).
\end{align}
The Loschmidt amplitude $G(t)$ for periodic boundary conditions can be simply written as \cite{karrasch2017dynamical,wu2020nonequilibrium},
\begin{align}
G(t)= \text{tr}\mathbf{T}^{N},
\end{align}
analogous to the partition function of the Ising model.
Here $\mathbf{T}$ is a $q \times q$ matrix with the elements, 
\begin{align}
\mathbf{T}_{m, n}=\frac{1}{q} e^{iJt2 \cos(2\pi(m-n)/q)},
\end{align}
and $m,n=0,1,...,q-1.$
The Loschmidt amplitude can also be written as \cite{karrasch2017dynamical,wu2020nonequilibrium},
\begin{align}
G(t)= \sum_{i=1}^{q}\Lambda_{i}^{N},
\label{Eq:generalGt}
\end{align}
in terms of eigenvalues $\Lambda_{i}$ of the matrix $\mathbf{T}$. Equation (\ref{Eq:generalGt}) allows one to analytically investigate the behaviors of the Loschmidt echo, which we will go into details later.

Let us introduce a time-evolved order parameter defined by,
\begin{align}
W(t) =& \frac{1}{N}\langle \psi(t) | \sum_{j} ( V_{j}+V^{\dagger}_{j} ) |\psi(t) \rangle,
\end{align}
using the potential operator $V_{j}$. If $V_{j}$ is written in terms of the $\{ \ket{m} \}$ as,  
\begin{align}
V_{j} = \sum_{m=0}^{q-1} \ket{m} \bra{m+1},
\end{align}
with a condition $\ket{q} \equiv \ket{0}$,
the order parameter $W(t)$ can be simply derived as (see Appendix \ref{App:A} for details),
\begin{align}
W(t) =& \frac{1}{N} \langle\psi_{0}|e^{-iJt\sum_{i}(U_{i+1}^{\dagger}U_{i} + U_{i}^{\dagger}U_{i+1})} \sum_{j}(V_{j}+V^{\dagger}_{j}) \nonumber \\
& \hspace{2cm} \times e^{iJt\sum_{i}(U_{i+1}^{\dagger}U_{i} + U_{i}^{\dagger}U_{i+1})} |\psi_{0}\rangle \\
=& \frac{1}{q^{3}}\sum_{\substack{m=0}}^{q-1}(\sum_{n=1}^{q-1}e^{-4iJt\sin(\frac{\theta}{2})\sin((m-n+\frac{1}{2})\theta)})^{2} + h.c,
\label{Eq:generalOP}
\end{align}
under periodic boundary conditions. We will use the order parameter in Eq.(\ref{Eq:generalOP}) to identify the DQPTs.

\subsection {Dynamics for $q \leq 4$}
Let us first consider the simplest case $q=2$, for which the matrix
\begin{align}
\mathbf{T} =\frac{1}{2}
\left( \begin{array}{ccc}
e^{2iJt} & e^{-2iJt} \\
e^{-2iJt} & e^{2iJt}
\end{array} \right).
\end{align}
The two eigenvalues of $\mathbf{T}$ are $\Lambda_{1} = \frac{1}{2} e^{-2iJt} (e^{4iJt}-1)$ and $\Lambda_{2}=\frac{1}{2} e^{-2iJt} (e^{4iJt}+1)$, respectively.
Therefore we have the Loschmidt amplitude
\begin{align}
G(t)=\left[ \frac{1}{2}e^{-2iJt}(e^{4iJt}-1) \right]^{N} + \left[ \frac{1}{2}e^{-2iJt}(e^{4iJt}+1) \right]^{N}.
\end{align}
The critical times for $q =2$ are given by,
\begin{align}
t_{cn} = \frac{\pi}{8J}(2n+1),
\end{align}
by finding the zeros (or minima) of the Loschmidt echo $L(t)= |G(t)|^{2}$ with $n \in \mathbb{N}$.
At the first critical point $t_{c1}=\frac{\pi}{8J}$, we obtain the rate function,
\begin{align}
r(t_{c1}) =& -\frac{1}{N} \ln L(t_{c1}) \nonumber \\
=& \ln2 - \frac{1}{N} \ln (1+i^{N})^{2},
\end{align}
which becomes $\ln2$ as the system size $N$ tends to infinity. This interesting result shows that the rate function $r(t_{c1})$ does not diverge but converges to a finite value in the thermodynamic limit.

\begin{figure}[tb]
\includegraphics[width=8.9cm]{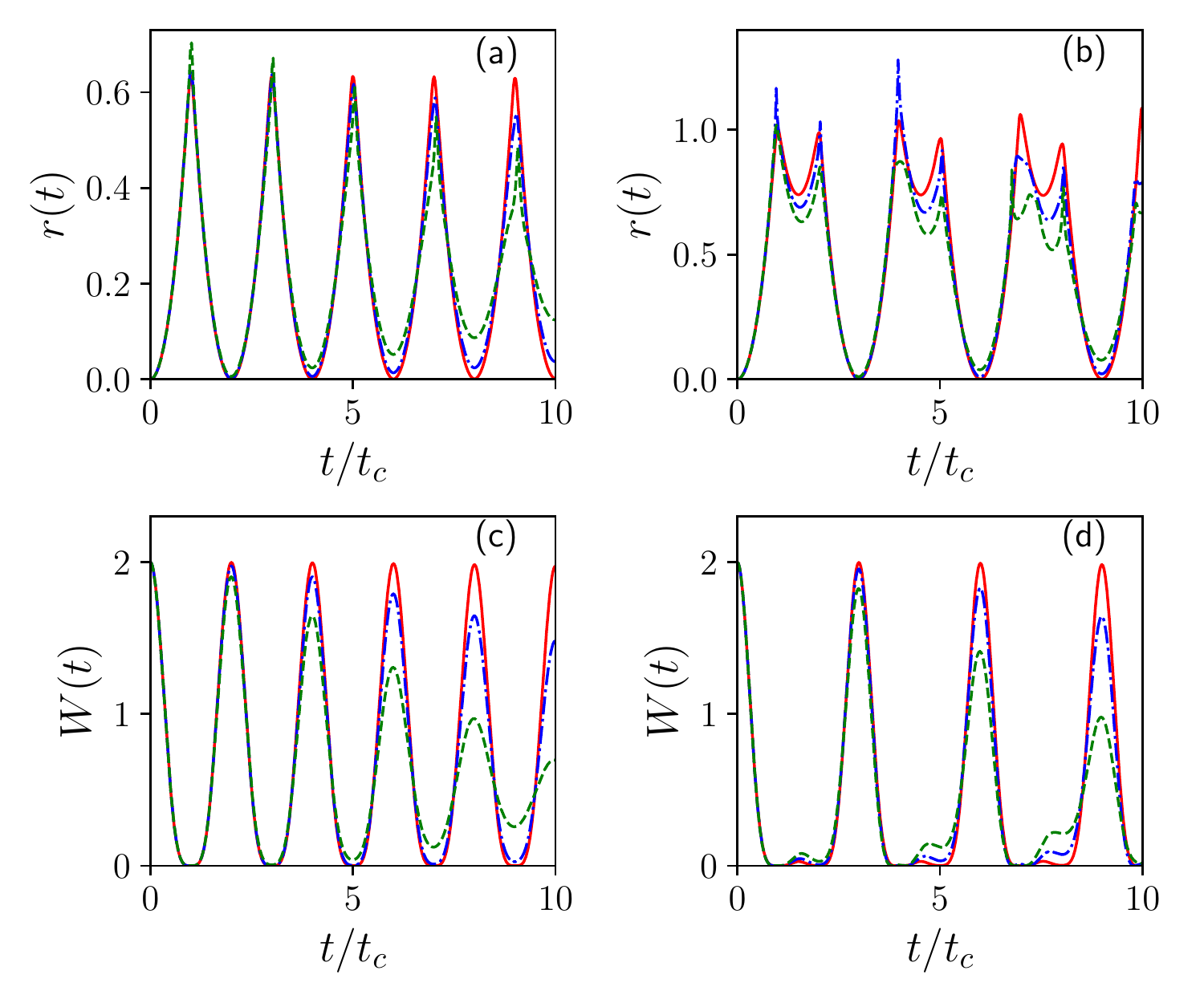}
\caption{Rate functions and order parameters. 
The dynamics is obtained for $J=1$ and $N = 24$ lattice sites by quenching the system from $h_{i} = \infty$ to $h_{f}=0.01$ (red solid line) , $h_{f}=0.05$ (blue dashed-dot line) and $h_{f}=0.1$ (green dashed line) 
with t-DMRG, respectively. 
(a) The rate functions $r(t)$ as the functions of $t/t_{c}$ for $q=2$ with $t_{c}=\pi/8$.
(c) The order parameters $W(t)$ as the functions of $t/t_{c}$ for $q=2$.
The corresponding data for $q=3$ with $t_{c}=2\pi/9$ are shown in (b) and (d), respectively.}
\label{RateDMRG234h}
\end{figure}

Let us continue to investigate whether the convergence of the rate function $r(t_{c1})$ in the thermodynamic limit persists for $q=3$ and $q=4$. Firstly, the matrix $\mathbf{T}$ for $q=3$ is given by
\begin{align}
\mathbf{T} =\frac{1}{3}
\left( \begin{array}{ccc}
e^{2iJt} & e^{-iJt} & e^{-iJt} \\
e^{-iJt} & e^{2iJt} & e^{-iJt} \\
e^{-iJt} & e^{-iJt} & e^{2iJt}
\end{array} \right).
\end{align}
The eigenvalues of $\mathbf{T}$ are $\Lambda_{1} = \Lambda_{2} = \frac{1}{3}e^{-iJt}(e^{3iJt}-1)$ and $\Lambda_{3} = \frac{1}{3}e^{-iJt}(e^{3iJt}+2)$, respectively.
We obtain the Loschmidt amplitude
\begin{align}
G(t)=2\left[ (\frac{1}{3} e^{-iJt}(e^{3iJt}-1) \right]^{N} + \left[ \frac{1}{3} e^{-iJt} (e^{3iJt}+2) \right]^{N}.
\end{align}
The critical times for $q=3$ are given by \cite{karrasch2017dynamical},
\begin{align}
t_{cn}^{1}=\frac{2\pi}{9J}(3n+1), \\
t_{cn}^{2}=\frac{2\pi}{9J}(3n+2).
\end{align}
The rate function is found to be $\ln3$ at the first critical time $t_{c1}=t_{c1}^{1}=\frac{2\pi}{9J}$ as the system size $N$ tends to infinity (see Appendix \ref{App:B} for details). 
Secondly, for $q=4$, the matrix $\mathbf{T}$ is given by
\begin{align}
\mathbf{T} =\frac{1}{4}
\left( \begin{array}{cccc}
e^{2iJt} & 1 & e^{-2iJt} & 1 \\
1 & e^{2iJt} & 1 & e^{-2iJt}\\
e^{-2iJt} & 1 & e^{2iJt} & 1 \\
1 & e^{-2iJt} & 1 & e^{2iJt}
\end{array} \right).
\end{align}
The eigenvalues of $\mathbf{T}$ are $\Lambda_{1} = \frac{1}{4} e^{-2iJt}(e^{2iJt}-1)^{2}$ , $\Lambda_{2} = \frac{1}{4}e^{-2iJt}(e^{2iJt}+1)^{2}$, 
$\Lambda_{3} = \Lambda_{4} = \frac{1}{4} e^{-2iJt}(e^{4iJt}-1)$, respectively.
Therefore we have
\begin{align}
G(t)=& \left[ \frac{1}{4} e^{-2iJt}(e^{2iJt}-1)^{2} \right]^{N} + \left[ \frac{1}{4}e^{-2iJt}(e^{2iJt}+1)^{2} \right]^{N} \notag \\
& + 2\left[ \frac{1}{4} e^{-2iJt}(e^{4iJt}-1) \right]^{N}.
\end{align}
From this, we find that the critical times for $q =4$ are given by
\begin{align}
t_{cn} = \frac{\pi}{4J}(2n+1).
\end{align} 
The rate function is found to be $\ln4$ at the first critical time $t_{c1}=\frac{\pi}{4J}$ as the system size $N$ tends to infinity (see Appendix \ref{App:B} for details). 
Therefore, we have proved that the rate function increases logarithmically with $q$ at the first critical time $t_{c1}$ for $q \leq 4$ (c.f. Fig.\ref{rateorderp234}(a)).
Moreover, analytical results show that the rate function $r(t)$ for $q=4$ is twice as big as that of $q=2$ (see Fig.\ref{rateorderp234}(a) and Appendix \ref{App:C} for details).

We have studied the rate function of the Loschmidt echo for an arbitrary finite system $N$ for $q \leq 4$. 
In the following, we will discuss the order parameter $W(t)$ of DQPTs, which is argued to zero at critical times that can be connected to the Loschmidt echo singularity for $q\leq4$. 
For $q=2$, the order parameter $W(t)$ is given by (see Appendix \ref{App:D} for details),
\begin{align}
W(t)=&\frac{1}{2^{3}}\sum_{m} \left( \sum_{n}e^{-4iJt\sin(\frac{\pi}{2})\sin((m-n+\frac{1}{2})\pi)} \right)^{2}+h.c \nonumber \\
=&2\cos^{2}(4Jt).
\end{align}
Likewise, we find that the order parameters $W(t)$ are equal to $\frac{2}{9}[(2\cos(3Jt)+1)^{2}]$ and $2\cos^{2}(2Jt))$ for $q=3$ and $q=4$, respectively. 
As shown in Fig.\ref{rateorderp234}(b), we find a one-to-one relationship between the Loschmidt echo singularity and the zeros of the order parameter (see Appendix \ref{App:D} for details).

In order to consider quantum fluctuations during the time evolution, we quench the system from $h_{i}=\infty$ to $h_{f}=0.01$, $h_{f}=0.05$ and $h_{f}=0.1$. 
The corresponding Loschmidt echoes and order parameters are computed by using the t-DMRG method for $N = 24$ lattice sites with periodic boundary conditions.
We find that DQPTs can survive in short time, where rate functions $r(t)$ of the Loschmidt echoes display kinks, 
which can also be characterized by the order parameters $W(t)$ as in the case of $h_{f}=0$ [c.f Fig.\ref{RateDMRG234h}].

\begin{figure}[tb]
\includegraphics[width=8.6cm]{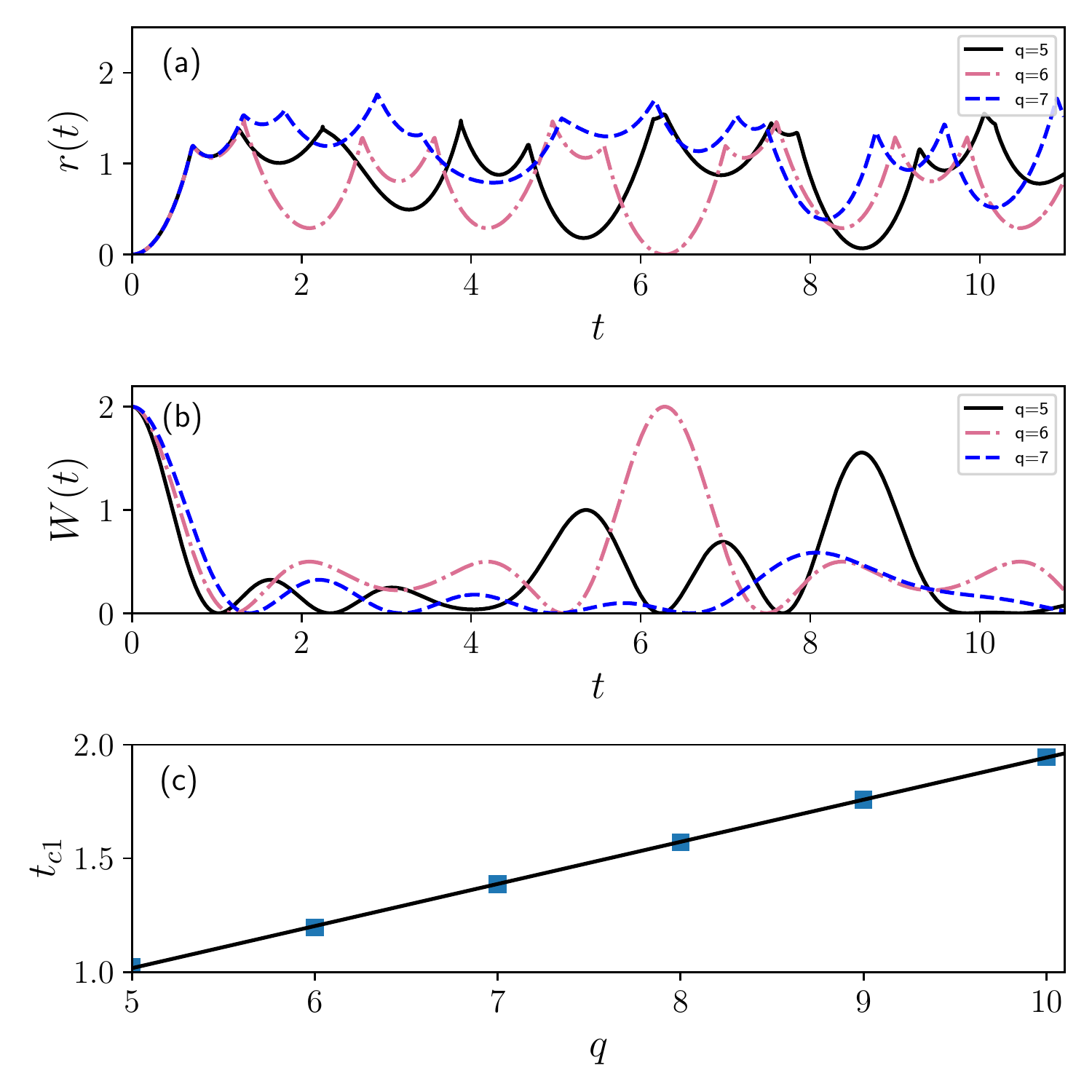}
\caption{Quench dynamics from $h_{i} = \infty$ to $h_{f}=0$ for $q > 4$.
(a) Rate functions $r(t)$ as the functions of $t$ for $q=5,6,7$ with $J=1$ and $N=100$ lattice sites, which tends to converge at the first cusp.
(b) Order parameters $W(t)$ with the same parameters as (a).
(c) The time $t_{c1}$ obtained from the first zeros of order parameters $W(t)$ with respect to the $q$. The blue square symbols are numerical results, and the black solid line is the fitting curve.}
\label{BigQuenchp567}
\end{figure}

\subsection {Dynamics for $q > 4$}
In the previous section, we have shown our main results of DQPTs for $q \leq 4$. It is fascinating to study DQPTs upon a quench across a critical point belonging to different universality classes.
For example, an exact mapping of DQPTs can be established for a quench across the Ising transition and a quench across a deconfined quantum critical point in a chain \cite{sun2020dynamical}. 
Since the $q$-state clock model exhibits BKT transitions for $q>4$, studying a quench dynamics for $q>4$ can shed a light on the nature of DQPTs upon a quench across a BKT transition 
and we will present the results in the following. 
The general results of the Loschmidt amplitude $G(t)$ in Eq.(\ref{Eq:generalGt}) and the order parameter $W(t)$ in Eq.(\ref{Eq:generalOP}) remain valid
for the cases $q > 4$. However, it is difficult to obtain the simple analytical formulas for $q >4$.
We will discuss the DQPTs based on the numerical results.

We find that the DQPTs persist for $q >4$, where rate functions $r(t)$ exhibit non-analytical behaviors associated with linear cusps [c.f. Fig.\ref{BigQuenchp567}(a)].
However, the Loschmidt echo for $q >4$ at the first critical time converges; this is in stark contrast with the logarithmical increase with $q$ for $q <4$ that we have found in the previous section.
Interestingly, the zeros of order parameters no longer correspond to the Loschmidt echo singularities [c.f. Fig.\ref{BigQuenchp567}(b)]; this observation indicates that the DQPTs for $q>4$ are beyond two-level dynamics. 
In addition, we find that the critical times $t_{c1}$ obtained from the first zeros of the order parameters $W(t)$ increase linearly with the $q$ as shown in Fig.\ref{BigQuenchp567}(c) for $q>4$. 
Our numerical calculations show that the DQPTs upon a big quench through a BKT transition can exhibit fundamentally different nature than that of a second-order phase transition. 
Our observation calls for a rigorous theoretical framework for understanding DQPTs across the BKT transition, which will be a topic of future studies.

\section {Conclusion}
\label{sec:conclusion}
In summary, we have studied the finite-size scaling laws of the Loschmidt echo in the quantum $q$-state clock model. 
We have shown that the short-time average rate function and the first minimum of the Loschmidt echo as a function of both time and coupling constant 
can serve as probes to detect equilibrium second-order phase transitions without knowing the accurate critical values in advance in the presence of discrete $Z_q$ symmetry.
The equilibrium correlation-length critical exponents $\nu$ obtained for $q \leq 4$ are consistent with the known results. 
It would be interesting to establish dynamical scaling laws of the Loschmidt echo for BKT transitions
to know whether the Loschmidt echo can characterize BKT transitions as the ground-state fidelity \cite{sun2019phase} in the future.

We have presented analytic results for the Loschmidt echo and the order parameter for any $q$, which have been used to understand DQPTs of the $q$-state clock model.
For $q\leq4$, we have shown that the Loschmidt echo singularity is connected to the zeros of the order parameter. 
In particular, the rate function is found to increase logarithmically with $q$ at the critical times.
Meanwhile, for $q>4$, the one-to-one correspondence between the Loschmidt echo and the order parameter no longer exists. 
The nature of DQPTs upon a quench across the BKT transition remains to be understood.

\begin{acknowledgments}
G.S. was supported by the NSFC under the Grants No. 11704186 and No. 11874220. 
W.-L.Y. is appreciative of support from the NSFC under the Grant No. 12174194, the startup fund (Grant No. 1008-YAH20006) of Nanjing University of Aeronautics and Astronautics, 
Top-notch Academic Programs Project of Jiangsu Higher Education Institutions, and stable support for basic institute research (Grant No.190101). 
M.-J. H. was supported by NSFC under the Grant No. 12050410258, the Startup Fund from Duke Kunshan University, and Innovation Program for Quantum Science and Technology 2021ZD0301602.
Numerical simulations were carried out on clusters of Nanjing University of Aeronautics and Astronautics.
\end{acknowledgments}

\bibliographystyle{apsrev4-1}
\bibliography{refclock}

\begin{widetext}
\appendix
\section{Derivation of order parameters}
\label{App:A}
In this section, we will show the details of the derivation of the order parameter $W(t)$, which is defined by,
\begin{align}
W(t) =& \frac{1}{N} \sum_{j} \langle V_{j}+V_{j}^{\dagger} \rangle \\
= & \frac{1}{N}  \sum_{j} \langle\psi(t) | (V_{j}+V^{\dagger}_{j}) |\psi(t) \rangle \\
=& \langle\psi_{0}|e^{-iJt\sum_{i}(U_{i+1}^{\dagger}U_{i} + U_{i}^{\dagger}U_{i+1})} (V_{1}+V^{\dagger}_{1}) 
   e^{iJt\sum_{i}(U_{i+1}^{\dagger}U_{i} + U_{i}^{\dagger}U_{i+1})} |\psi_{0}\rangle \\
=&\langle \psi_{0}| e^{-iJt[(U_{2}^{\dagger}+U_{N}^{\dagger})U_{1} + U_{1}^{\dagger}(U_{2}+U_{N})]} (V_{1}+V^{\dagger}_{1}) 
  e^{iJt[(U_{2}^{\dagger}+U_{N}^{\dagger})U_{1} + U_{1}^{\dagger}(U_{2}+U_{N})]}|\psi_{0}\rangle
\end{align}
Remember the operator $V_j$ is,
\begin{align}
V_j = \sum_{m=0}^{q-1} \left( |m\rangle \langle m+1| \right ).
\end{align}
we have,
\begin{align}
W(t)=&\langle \psi_{0}| e^{-iJt[(U_{2}^{\dagger}+U_{N}^{\dagger})U_{1} + U_{1}^{\dagger}(U_{2}+U_{N})]}
\sum_{\substack{m=0}}^{q-1}(|m\rangle\langle m+1|)_{1}
e^{iJt[(U_{2}^{\dagger}+U_{N}^{\dagger})U_{1} + U_{1}^{\dagger}(U_{2}+U_{N})]}|\psi_{0}\rangle + h.c, \\
=&\sum_{\substack{m=0}}^{q-1}\langle \psi_{0}| e^{-iJt[e^{im\theta}(U_{2}^{\dagger}+U_{N}^{\dagger}) + e^{-im\theta}(U_{2}+U_{N})]}
(|m\rangle\langle m+1|)_{1}e^{iJt[e^{i(m+1)\theta}(U_{2}^{\dagger}+U_{N}^{\dagger})+e^{-i(m+1)\theta}(U_{2}+U_{N})]}|\psi_{0}\rangle + h.c, \\
=&\sum_{\substack{m=0}}^{q-1}\langle \psi_{0}|
e^{iJt[(e^{i(m+1)\theta}-e^{im\theta})(U_{2}^{\dagger}+U_{N}^{\dagger})+(e^{-i(m+1)\theta}-e^{-im\theta})(U_{2}+U_{N})]}
(|m\rangle\langle m+1|)_{1}|\psi_{0}\rangle + h.c.
\end{align}
Because,
\begin{align}
|\psi_{0}\rangle = \bigotimes\limits^{N}_{j=1} (\frac{1}{\sqrt{q}}\sum_{n}|n\rangle)_{j}
\end{align}
and,
\begin{align}
\langle m|\frac{1}{\sqrt{q}}\sum_{n}|n\rangle = \frac{1}{\sqrt{q}}
\end{align}
We have,
\begin{align}
W(t)=&\frac{1}{q}\sum_{\substack{m=0}}^{q-1}\langle \psi_{0}|e^{iJt[(e^{i(m+1)\theta}-e^{im\theta})U_{2}^{\dagger}+(e^{-i(m+1)\theta}-e^{-im\theta})U_{2}]}
           e^{iJt[(e^{i(m+1)\theta}-e^{im\theta})U_{N}^{\dagger}+(e^{-i(m+1)\theta}-e^{-im\theta})U_{N}]}|\psi_{0}\rangle + h.c \\
=&\frac{1}{q}\sum_{\substack{m=0}}^{q-1}(\frac{1}{q}\sum_{n,n^{\prime},n^{\prime \prime}} {_{2}\langle} n| e^{iJt[(e^{i(m+1)\theta}-e^{im\theta})U_{2}^{\dagger}]}
(|n^{\prime}\rangle\langle n^{\prime}|)_{2} e^{iJt[(e^{-i(m+1)\theta}-e^{-im\theta})U_{2}]}|n^{\prime \prime}\rangle_{2})^{2} + h.c \\
=&\frac{1}{q}\sum_{m=0}^{q-1}\frac{1}{q^{2}} (\sum_{n} e^{iJt[(e^{i(m+1)\theta}-e^{im\theta})e^{-in\theta}] e^{iJt[(e^{-i(m+1)\theta}-e^{-im\theta})e^{in\theta}]}})^{2} + h.c \\
=&\frac{1}{q^{3}}\sum_{\substack{m=0}}^{q-1}(\sum_{n}e^{iJt[\substack{(e^{i\theta}-1)e^{i(m-n)\theta} +(e^{-i\theta}-1)e^{i(n-m)\theta}}]})^{2} + h.c \\
=&\frac{1}{q^{3}}\sum_{\substack{m=0}}^{q-1}(\sum_{n}e^{iJt[\substack{e^{\frac{i\theta}{2}}(e^{\frac{i\theta}{2}}-e^{-\frac{i\theta}{2}})e^{i(m-n)\theta}
 +e^{-\frac{i\theta}{2}}(e^{-\frac{i\theta}{2}}-e^{\frac{i\theta}{2}})e^{i(n-m)\theta}}]})^{2}+ h.c \\
=&\frac{1}{q^{3}}\sum_{\substack{m=0}}^{q-1}(\sum_{n}e^{iJt[\substack{(e^{\frac{i\theta}{2}}-e^{-\frac{i\theta}{2}})(e^{(m-n+\frac{1}{2})i\theta}-e^{-(m-n+\frac{1}{2})i\theta})}]})^{2}+h.c \\
=&\frac{1}{q^{3}}\sum_{\substack{m=0}}^{q-1}(\sum_{n}e^{iJt[2i\sin(\frac{\theta}{2}) \cdot 2i\sin((m-n+\frac{1}{2})\theta)]})^{2}+h.c \\
=&\frac{1}{q^{3}}\sum_{\substack{m=0}}^{q-1}(\sum_{n}e^{-4iJt\sin(\frac{\theta}{2})\sin((m-n+\frac{1}{2})\theta)})^{2}+h.c
\end{align}
We arrive at the general analytic solution of the order parameter $W(t)$.

\section{Derivation of the rate function at first critical times}
\label{App:B}
In this section, we present the details for the rate function at the first critical time $t_c$.
When $q=2$, Loschmidt amplitude
\begin{align}
G(t)=(\frac{1}{2}e^{-2iJt}(e^{4iJt}-1))^{N}+(\frac{1}{2}e^{-2iJt}(e^{4iJt}+1))^{N}.
\end{align}
Place the first critical time $t=t_{c1}=\frac{\pi}{8J}$ into it, we have
\begin{align}
G(t_{c1}) = 2^{-\frac{N}{2}}(1+i^{N}).
\end{align}
The rate function $r(t)$ is written by,
\begin{align}
r(t_{c1}) =& -\frac{1}{N} \ln L(t_{c1}), \nonumber \\
      =& \ln2 - \frac{2}{N} \ln |1+i^{N}|.
\end{align}
If the size of the system $N$ tends to infinity, the rate function turns into $\ln2$.

When $q=3$, The matrix $\mathbf{T}$ is,
\begin{align}
\mathbf{T} =\frac{1}{3}
\left( \begin{array}{ccc}
e^{2iJt} & e^{-iJt} & e^{-iJt} \\
e^{-iJt} & e^{2iJt} & e^{-iJt} \\
e^{-iJt} & e^{-iJt} & e^{2iJt}
\end{array} \right)
\end{align}
The eigenvalues of $\mathbf{T}$ are $\Lambda_{1} = \Lambda_{2} = \frac{1}{3}e^{-iJt}(e^{3iJt}-1)$ and $\frac{1}{3}e^{-iJt}(e^{3iJt}+2)$, respectively. Therefore we have,
\begin{align}
G(t)=2[\frac{1}{3} e^{-iJt}(e^{3iJt}-1)]^{N}+[\frac{1}{3} e^{-iJt} (e^{3iJt}+2)]^{N}
\end{align}
When placing the first critical time $t=t_{c}=\frac{2\pi}{9J}$ into $G(t)$, we have the following simple form,
\begin{align}
G(t_{c1}) = 3^{-N}e^{-i2N\pi/9}[2(e^{i2\pi/3}-1)^N + (e^{i2\pi/3}+2)^N]
\end{align}
Then the rate function is given by,
\begin{align}
r(t_{c1}) =& -\frac{1}{N} \ln L(t_{c1}) \nonumber \\
      =& \ln3 - \frac{1}{N}\ln|5+4(-1)^N \cos(N\pi/3)|
\end{align}
It is not difficult to see if the size of the system $N$ verge to infinity, the rate function turn into $\ln3$.

Similarly, the matrix $\mathbf{T}$ for $q=4$ is,
\begin{align}
\mathbf{T} =\frac{1}{4}
\left( \begin{array}{cccc}
e^{2iJt} & 1 & e^{-2iJt} & 1 \\
1 & e^{2iJt} & 1 & e^{-2iJt}\\
e^{-2iJt} & 1 & e^{2iJt} & 1 \\
1 & e^{-2iJt} & 1 & e^{2iJt}
\end{array} \right)
\end{align}
The eigenvalues of $\mathbf{T}$ are $\Lambda_{1} = \frac{1}{4J} e^{-2iJt}(e^{2iJt}-1)^{2}$ , $\Lambda_{2} = \frac{1}{4}e^{-2iJt}(e^{2iJt}+1)^{2}$, $\Lambda_{3} = \Lambda_{4} = \frac{1}{4} e^{-2iJt}(e^{4iJt}-1)$. 
Therefore we have,
\begin{align}
G(t) = [\frac{1}{4} e^{-2iJt} (e^{2iJt}-1)^{2}]^{N}+[\frac{1}{4} e^{-2iJt} (e^{2iJt}+1)^{2}]^{N} + 2[\frac{1}{4} e^{-2iJt} (e^{4iJt}-1)]^{N}
\end{align}
When the critical time $t=t_{c1}=\frac{\pi}{4J}$ is placed into $G(t)$, we have,
\begin{align}
G(t_{c1}) = 2^{-N}[1+(-1)^{N}+2i^{N}]
\end{align}
Then the rate function for $q=4$ is obtained by,
\begin{align}
r(t_{c1})  =& -\frac{1}{N} \ln L(t_{c1}) \nonumber \\
       =& \ln4 - \frac{2}{N} \ln|1+(-1)^{N}+2 i^{N}|
\end{align}
Obviously, If the size of the system $N$ tends to infinity, the rate function turn into $\ln4$.

\section{Relationship of rate functions for $q=2$ and $q=4$}
\label{App:C}
In the section, we will go into detail about the relationship between the rate function of $q=2$ and $q=4$.
We prove that the rate function $r(t)$ for $q=4$ is twice as much as that of $q=2$ as the function of $t/t_{c1}$.
For $q=2$, we have the Loschmidt amplitude,
\begin{align}
G(t)=[i\sin2Jt]^{N}+[\cos2Jt]^{N}
\end{align} 
As there are four cases based on $i^N$, $G(t)$ have four forms,
\begin{align}
G(t) =& i(\sin2Jt)^{N}+(\cos2Jt)^{N},  \hspace{2.02cm} (N=4n+1) \\
G(t) =& -(\sin2Jt)^{N}+(\cos2Jt)^{N},  \hspace{1.7cm} (N=4n+2) \\
G(t) =& -i(\sin2Jt)^{N}+(\cos2Jt)^{N}, \hspace{1.6cm} (N=4n+3) \\
G(t) =& (\sin2Jt)^{N}+(\cos2Jt)^{N},   \hspace{2.18cm} (N=4n) 
\end{align}
The Loschmidt echo has the following forms:
\begin{align}
L(t)=& (\sin2Jt)^{2N}+(\cos2Jt)^{2N},   \hspace{2.0cm} (N=4n+1) \\
L(t)=& \left[(\sin2Jt)^{N}-(\cos2Jt)^{N} \right]^{2},   \hspace{1.7cm} (N=4n+2) \\
L(t)=& (\sin2Jt)^{2N}+(\cos2Jt)^{2N},   \hspace{2.0cm} (N=4n+3) \\
L(t)=& \left[(\sin2Jt)^{N}+(\cos2Jt)^{N} \right]^{2},   \hspace{1.7cm} (N=4n) 
\end{align}
For $q=4$, as $\Lambda_{1}=-\sin^{2}Jt$, $\Lambda_{2}=\cos^{2}Jt$, $\Lambda_{3}=\Lambda_{4}=\frac{i}{2}\sin2Jt$,
we have the Loschmidt amplitude,
\begin{align}
G(t)=[-\sin^{2}Jt]^{N}+[\cos^{2}Jt]^{N}+2[\frac{i}{2}\sin2Jt]^{N},
\end{align}
which has the following forms,
\begin{align}
G(t)=& [-\sin^{2}Jt]^{N}+[\cos^{2}Jt]^{N}+2i[\frac{1}{2}\sin2Jt]^{N} \hspace{2.0cm} (N=4n+1) \\
G(t)=& [-\sin^{2}Jt]^{N}+[\cos^{2}Jt]^{N}-2[\frac{1}{2}\sin2Jt]^{N} \hspace{2.1cm} (N=4n+2) \\
G(t)=& [-\sin^{2}Jt]^{N}+[\cos^{2}Jt]^{N}-2i[\frac{1}{2}\sin2Jt]^{N} \hspace{2.0cm} (N=4n+3) \\
G(t)=& [-\sin^{2}Jt]^{N}+[\cos^{2}Jt]^{N}+2[\frac{1}{2}\sin2Jt]^{N} \hspace{2.18cm} (N=4n)
\end{align}
Consequently, the Loschmidt echo has the following forms:

When $N=4n+1$, it is,
\begin{align}
L(t)=&[(-\sin^{2}Jt)^{N}+(\cos^{2}Jt)^{N}+2i(\frac{1}{2}\sin2Jt)^{N}] \cdot [(-\sin^{2}Jt)^{N}+(\cos^{2}Jt)^{N}-2i(\frac{1}{2}\sin2Jt)^{N}] \\
=& (\sin Jt)^{4N} + 2(\sin Jt\cos Jt)^{2N}+(\cos Jt)^{4N} \\
=& [(\sin Jt)^{2N}+(\cos Jt)^{2N}]^{2}
\end{align}
When $N=4n+2$, it is,
\begin{align}
L(t)=&[(-\sin^{2}Jt)^{N}+(\cos^{2}Jt)^{N}-2(\frac{1}{2}\sin2Jt)^{N}] \cdot [(-\sin^{2}Jt)^{N}+(\cos^{2}Jt)^{N}-2(\frac{1}{2}\sin2Jt)^{N}] \\
=& [(\sin^{2}Jt)^{N}+(\cos^{2}Jt)^{N}-2(\frac{1}{2}\sin2Jt)^{N}] \cdot [(\sin^{2}Jt)^{N}+(\cos^{2}Jt)^{N}-2(\frac{1}{2}\sin2Jt)^{N}] \\
=&(\sin Jt)^{4N}+2(\sin Jt\cos Jt)^{2N}+(\cos Jt)^{4N} - 4(\frac{1}{2}\sin^{2}Jt\sin 2Jt)^{N}
  -4(\frac{1}{2}\cos^{2}Jt\sin 2Jt)^{N} + 4(\sin t\cos t)^{2N} \\
=&[(\sin Jt)^{2N}+(\cos Jt)^{2N}]^{2}-4(\sin Jt\cos Jt)^{N}[(\sin Jt)^{2N}+(\cos Jt)^{2N}]+4(\sin t \cos t)^{2N} \\
=&[(\sin Jt)^{2N}-2(\sin t \cos t)^{N}+(\cos Jt)^{2N}]^{2} \\
=&[(\sin Jt)^{N}-(\cos Jt)^{N}]^{4}
\end{align}
When $N=4n+3$, it is,
\begin{align}
L(t)=& [(-\sin^{2}Jt)^{N}+(\cos^{2}Jt)^{N}-2i(\frac{1}{2}\sin2Jt)^{N}] \cdot [(-\sin^{2}Jt)^{N}+(\cos^{2}Jt)^{N}+2i(\frac{1}{2}\sin2Jt)^{N}] \\
=& (\sin Jt)^{4N}+2(\sin Jt\cos Jt)^{2N}+(\cos Jt)^{4N} \\
=& [(\sin Jt)^{2N}+(\cos Jt)^{2N}]^{2}
\end{align}
When $N=4n$, it is,
\begin{align}
L(t)=&[(-\sin^{2}Jt)^{N}+(\cos^{2}Jt)^{N}+2(\frac{1}{2}\sin2Jt)^{N}] \cdot [(-\sin^{2}Jt)^{N}+(\cos^{2}Jt)^{N}+2(\frac{1}{2}\sin2Jt)^{N}] \\
=& [(\sin^{2}Jt)^{N}+(\cos^{2}Jt)^{N}+2(\frac{1}{2}\sin2Jt)^{N}] \cdot [(\sin^{2}Jt)^{N}+(\cos^{2}Jt)^{N}+2(\frac{1}{2}\sin2Jt)^{N}] \\
=&[(\sin Jt)^{2N}+2(\sin t \cos t)^{N}+(\cos Jt)^{2N}]^{2} \\
=&[(\sin Jt)^{N}+(\cos Jt)^{N}]^{4}
\end{align}
Remember the critical times are $t_{cn} = \frac{\pi}{8J}(2n+1)$ and $t_{cn} = \frac{\pi}{4J}(2n+1)$ for $q=2$ and $q=4$. 
It is easy to find that the Loschmidt echo for $q=4$ is merely the square of the Loschmidt echo for $q=2$ as the function of $t/t_{c1}$.  
In other words, the rate function of the Loschmidt echo for $q=4$ are twice as big as that for $q=2$ as the function of $t/t_c$.

\section{The order parameters of $q=2,3,4$}
\label{App:D}
When $q=2$, $\theta=\frac{2\pi}{2}=\pi$, we can obtain the order parameter:
\begin{align}
W(t)=& \frac{1}{N} \sum_{j} \langle V_{j}+V_{j}^{\dagger} \rangle \\
=& \frac{1}{2^{3}}\sum_{\substack{m=0}}^{1}(\sum_{n=0}^{1}e^{-4iJt\sin(\frac{\pi}{2})\sin((m-n+\frac{1}{2})\pi)})^{2}+h.c \\
=& \frac{1}{8}\sum_{\substack{m=0}}^{1}(e^{-4iJt\sin(\frac{\pi}{2})\sin((m+\frac{1}{2})\pi)} + e^{-4iJt\sin(\frac{\pi}{2})\sin((m-\frac{1}{2})\pi)})^{2}+h.c \\
=& 2\times\frac{1}{8}(e^{-4iJt}+e^{-4iJt})^{2}+ h.c \\
=& \cos^{2}(4Jt)+ h.c \\
=& 2\cos^{2}(4Jt)
\end{align}
When $q=3$, $\theta=\frac{2\pi}{3}$, the order parameter is,
\begin{align}
W(t)=& \frac{1}{N} \sum_{j} \langle V_{j}+V_{j}^{\dagger} \rangle \\
=& \frac{1}{3^{3}}\sum_{m=0}^{2}(\sum_{n=0}^{2}e^{-4iJt\sin(\frac{\pi}{3})\sin((m-n+\frac{1}{2})\frac{2\pi}{3})})^{2}+h.c \\
=& \frac{1}{27}\sum_{m=0}^{2}(e^{-4iJt\sin(\frac{\pi}{3})\sin((m+\frac{1}{2})\frac{2\pi}{3})}+e^{-4iJt\sin(\frac{\pi}{3})\sin((m-\frac{1}{2})\frac{2\pi}{3})}+e^{-4iJt\sin(\frac{\pi}{3})\sin((m-\frac{3}{2})\frac{2\pi}{3})})^{2}+h.c \\
=& \frac{1}{27}[3(e^{-3iJt}+e^{3iJt}+1)^{2}]+ h.c \\
=& \frac{2}{9}[(2\cos(3Jt)+1)^{2}]
\end{align}

When $q=4$, $\theta=\frac{2\pi}{4}=\frac{\pi}{2}$, the order parameter is,

\begin{align}
W(t)=& \frac{1}{N} \sum_{j} \langle V_{j}+V_{j}^{\dagger} \rangle \\
=& \frac{1}{4^{3}}\sum_{m=0}^{3}(\sum_{n=0}^{3}e^{-4iJt\sin(\frac{\pi}{4})\sin((m-n+\frac{1}{2})\frac{\pi}{2})})^{2}+h.c \\
=& \frac{1}{64}\sum_{m=0}^{3}(e^{-4iJt\sin(\frac{\pi}{4})\sin((m+\frac{1}{2})\frac{\pi}{2})} + e^{-4iJt\sin(\frac{\pi}{4})\sin((m-\frac{1}{2})\frac{\pi}{2})} \nonumber \\
   & \hspace{2.0cm} + e^{-4iJt\sin(\frac{\pi}{4})\sin((m-\frac{3}{2})\frac{\pi}{2})} + e^{-4iJt\sin(\frac{\pi}{4})\sin((m-\frac{5}{2})\frac{\pi}{2})})^{2}+ h.c \\
=&\frac{1}{64}[4(e^{-2iJt}+e^{2iJt}+e^{2iJt}+e^{-2iJt})^{2}]+ h.c \\
=&\frac{1}{64}[4(4\cos(2Jt))^{2}]+ h.c \\
=&2\cos^{2}(2Jt)
\end{align}
We derive the simple forms of the order parameter for $q=2,3,4$.

\end{widetext}

\end{document}